\documentclass[aps,prl,twocolumn,superscriptaddress,showpacs]{revtex4-1}
\usepackage{amsmath,amssymb,wasysym,graphicx}
\usepackage[utf8]{inputenc}
\usepackage[T1]{fontenc}
\usepackage{xcolor}

\IfFileExists{newtxtext.sty}
{\usepackage{newtxtext,newtxmath}}
{\IfFileExists{stix.sty}
	{\usepackage{stix}}
	{\IfFileExists{mathptmx.sty}
		{\usepackage{mathptmx}}{} } }

\usepackage{textcomp}

\usepackage{bm}

\IfFileExists{siunitx.sty}{\usepackage{booktabs,siunitx}}{}

\usepackage{color}
\definecolor{LinkColor}{rgb}{0.256,0.439,0.588}
\usepackage{hyperref}
\hypersetup{
	pdfauthor={good guys},
	pdftitle={good title},
	colorlinks=true,
	citecolor=LinkColor,
	linkcolor=LinkColor,
	urlcolor=LinkColor
}

\def\normOrd#1{\mathop{:}\nolimits\!#1\!\mathop{:}\nolimits}
\newcommand{\beq} {\begin{equation}}
\newcommand{\eeq} {\end{equation}}
\newcommand{\bea} {\begin{eqnarray}}
\newcommand{\eea} {\end{eqnarray}}
\newcommand{\be} {\begin{equation}}
\newcommand{\ee} {\end{equation}}
\renewcommand{\(}{\left(}
\renewcommand{\)}{\right)}
\renewcommand{\[}{\left[}
\renewcommand{\]}{\right]}

\DeclareMathOperator{\Tr}{Tr}

\newcommand{\ket}[1]{\left|#1\right>}
\newcommand{\bra}[1]{\left<#1\right|}

\def\Eq#1{Eq.~(\ref{#1})}
\def\Fig#1{Fig.~\ref{#1}}
\def\avg#1{\left\langle#1\right\rangle}

\begin{document}

\title{Universal entanglement spectrum in gapless symmetry protected topological states}

\author{Xue-Jia Yu}
\altaffiliation{The first two authors contributed equally.}
\affiliation{Department of Physics, Fuzhou University, Fuzhou 350116, Fujian, China}
\affiliation{Fujian Key Laboratory of Quantum Information and Quantum Optics,
College of Physics and Information Engineering,
Fuzhou University, Fuzhou, Fujian 350108, China}

\author{Sheng Yang}
\altaffiliation{The first two authors contributed equally.}
\affiliation{Institute for Advanced Study in Physics and School of Physics, Zhejiang University, Hangzhou 310058, China}

\author{Hai-Qing Lin}
\affiliation{Institute for Advanced Study in Physics and School of Physics, Zhejiang University, Hangzhou 310058, China}

\author{Shao-Kai Jian}
\email{sjian@tulane.edu}
\affiliation{Department of Physics and Engineering Physics, Tulane University, New Orleans, Louisiana, 70118, USA}

\begin{abstract}
Quantum entanglement marks a definitive feature of topological states.
However, the entanglement spectrum remains insufficiently explored for topological states without a bulk energy gap. 
Using a combination of field theory and numerical techniques, we accurately calculate and analyze the entanglement spectrum of gapless symmetry protected topological states in one dimension. 
We highlight that the universal entanglement spectrum not only encodes the nontrivial edge degeneracy, generalizing the Li-Haldane conjecture to gapless topological states, but also contains the operator content of the underlying boundary conformal field theory. 
This implies that the bulk wave function can act as a fingerprint of both quantum criticality and topology in gapless symmetry protected topological states.
We also identify a symmetry enriched conformal boundary condition that goes beyond the conventional conformal boundary condition.

\end{abstract}

\maketitle

\emph{Introduction.}---Topological phases are novel many-body states featuring nonlocal order parameters and unusual entanglement properties.
It is well understood that the quantum entanglement structure is necessary to describe these topological phases, since they fail to be distinguished by local observable in the bulk.
For instance, as noticed in the famous Li-Haldane conjecture~\cite{li2008prl}, the bulk entanglement spectrum (ES) encodes the information on the boundary Hamiltonian. 
More explicitly, this conjecture states that the bulk low-lying ES is in one-to-one correspondence with the universal part of the many-body energy spectrum at the boundary of the system, which indicates that the bulk ground state wave function can capture boundary universal information, such as edge mode degeneracy of the gapped topological phase~\cite{chandran2011prb,yan2023unlocking,liu2023probing,Song_2023,li2023relevant,chen2003prb}.

Symmetry protected topological phases (SPT)~\cite{wen2017rmp,wen2019choreographed,gu2009prb,chen2012symmetry,yu2022emergence}, as one subclass of the topological phases, refers to the topological states that are only nontrivial under certain global symmetry.
While the bulk of SPTs is gapped, nontrivial gapless states emerge at the boundary. 
Despite the crucial role of the bulk gap in defining topological phases, recent research~\cite{keselman2015prb,cheng2011prb,scaffidi2017prx,fidkowski2011prb,kestner2011prb,iemini2015prl,lang2015prb,ruhman2017prb,JIANG2018753,keselman2018prb,yu2023pre,scaffidi2017prx,verresen2021prx,yu2022prl,parker2018prb,li2023intrinsicallypurely,li2023decorated,huang2023topological,wen2023prb,wen2023classification} has revealed that many key features of topological physics persist in the gapless case, even in the presence of the non-trivial coupling between the topological edge modes and the critical bulk modes. 
This extension is termed gapless symmetry-protected topological phases or symmetry enriched quantum critical points (QCPs)~\cite{verresen2021prx,verresen2018prl,verresen2020topology,liu2021prb,duque2021prb,jones2019asymptotic,yu2022prl,parker2018prb,yescipost2022,nathananscipost2023,nathanan_scipost2023,thorngren2021prb,chang2022absence,verresen2021quotient,prembabu2022boundary,umberto_scipost2021,li2023intrinsicallypurely,li2023decorated,li2023noninvertible,yang2023prb,wang2023stability,ys2023prb,huang2023topological,wen2023prb,wen2023classification,mondal2023prb,hidaka2022prb,florescalderón2023topological}, which we summarize under the name of gapless symmetry protected topological states (gSPTs). 
This development has led to the discovery of new critical points and phases in 1+1D with unusual string operators that imply symmetry-protected topological edge modes, classified by conformal boundary conditions~\cite{yu2022prl,parker2018prb}.

In the context of gapless topological states, where the bulk is at a symmetry enriched quantum critical point or a gapless symmetry protected phase, the question of how universal the Li-Haldane conjecture remains an interesting open question. 
We note that these gapless topological states not only host topological protected edge modes, but also have bulk critical fluctuations described by a conformal field theory (CFT)~\cite{sachdev2023quantum,sachdev_2011,cardy1996scaling,francesco2012conformal,ginsparg1988applied,scaffidi2017prx,verresen2021prx,yu2022prl,parker2018prb,li2023intrinsicallypurely,li2023decorated,huang2023topological,wen2023prb,wen2023classification,yu2022prb,cardy2004charge,cardy2009charge}.
The ES in CFTs has been extensively studied~\cite{li2008prl,calabrese2008pra,pollmann2010entanglement,qi2012prl,chandran2014prl,poilblanc2010prl,LAFLORENCIE20161,yao2010prl,fidkowski2010prl,zhou2023reviving,li2023numerical}, which shows that it contains universal information that goes beyond the entanglement entropy.
With powerful conformal invariance in two dimensions, the ES in a 1+1D CFT in various geometries can be exactly mapped to the energy spectrum with an open boundary condition~\cite{lauchli2013operator,Ohmori_2015,Cardy_2016,swingle2012prb}, aligning with the operator content of the underlying boundary CFT~\cite{CARDY1984514,CARDY1986200,CARDY1989581}.
Hence, this raises an intriguing question: 
On one hand, ES reveals the information on nontrivial boundary states for the topological phases according to the Li-Haldane conjecture. 
On the other hand, it also contains the operator content in a boundary CFT prescribed presumably by the entanglement cut.
What is the interplay between these two interesting phenomena in 1+1D gapless topological states? 
Can we extract the topology and boundary CFT information solely from the bulk wave function according to ES?

In this Letter, we study different families of quantum spin chains that exhibit different types of the gapless symmetry protected topological states. 
Each family contains symmetry protected topological edge modes that are described by the corresponding boundary CFT. 
By examining their ES and energy spectrum, we show that the bulk ES is in one-to-one correspondence with the energy spectrum at the edge of the system, which means that the ES contains the information of both the topological edge state and the corresponding boundary CFT operator content. 
Additionally, thanks to conformal symmetry present in the gSPTs, the universal spectrum correspondence can be understood theoretically, thus establishing a solid bulk-boundary correspondence in 1+1D gSPTs. 
We also identify a symmetry enriched conformal boundary condition in the free boson CFT beyond the conventional Dirichlet boundary condition.

\emph{ES of gapped SPT.}---The ES consists of the eigenvalues of the modular or entanglement Hamiltonian (EH) $\Tilde{H}_{A}$, which is related to the reduced density matrix $\rho_{A}$ of the subsystem $A$ by
\begin{equation}
    \label{E_rho}
    \rho_{A} = \text{Tr}_{B}\ket{\Psi}\!\bra{\Psi} = \sum_{\alpha} e^{-\ln\lambda_{\alpha}} \ket{\Psi_{\alpha}^{A}}\bra{\Psi_{\alpha}^{A}} = e^{-\Tilde{H}_{A}}\,.
\end{equation}
Here, $\ket{\Psi}$ is the ground state wave function of the Hamiltonian, $\lambda_{\alpha}$ is the eigenvalue of $\rho_{A}$.
In our study of 1D quantum chain, $A=\{1,2,...L/2\}$ and $B=\{L/2+1,...L\}$ represent a spatial bipartition of the whole chain. 
The boundary points between $A$ and $B$ (more generally, the boundary between $A$ and $B$) are called the entangling surface (or entanglement cut).  

\begin{figure}[tb]
\includegraphics[width=0.9\linewidth]{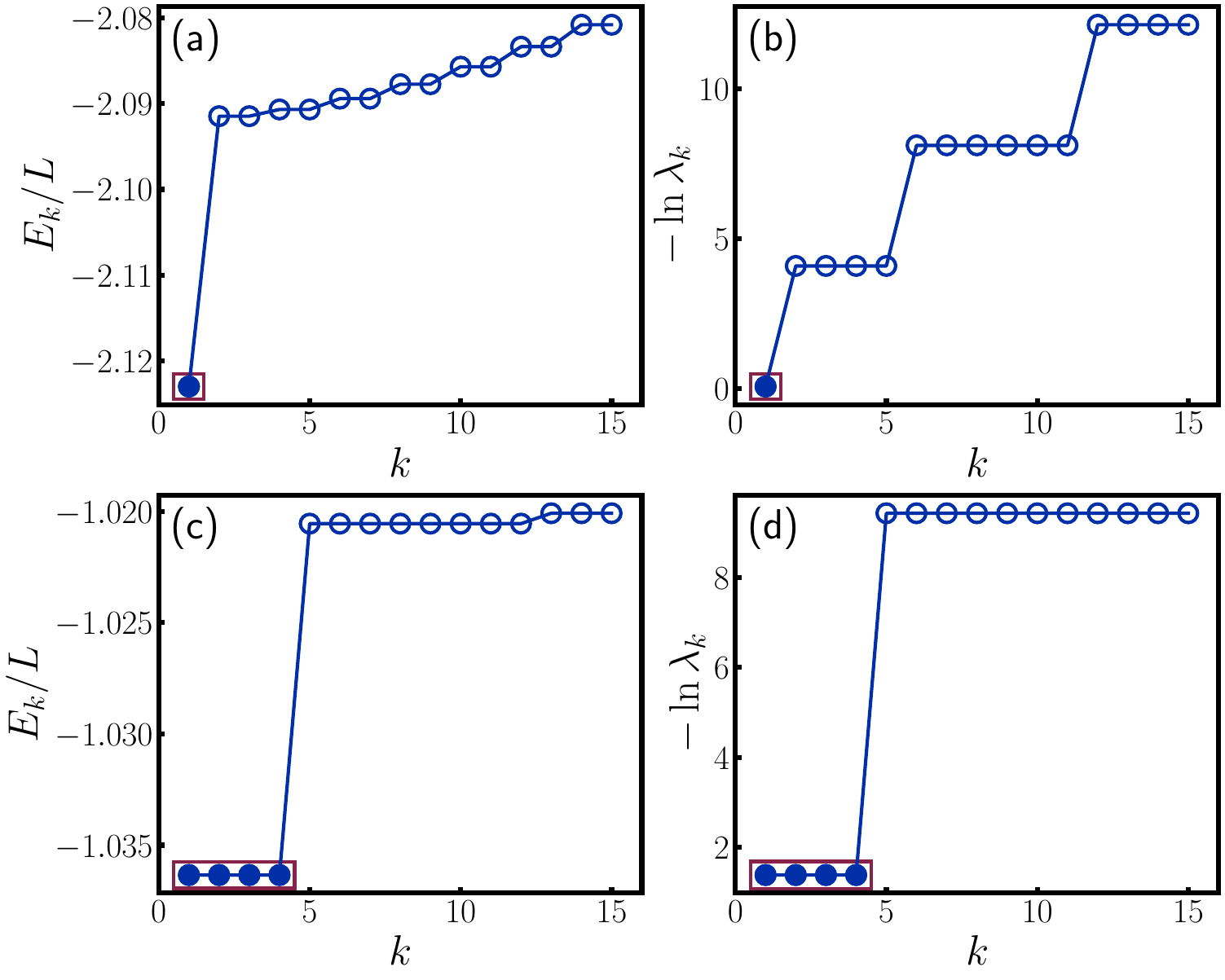}
\caption{(Color online) Probing Li-Haldane conjecture in (1+1)D cluster SPT model. OBC energy spectrum and PBC bulk ES for (a-b) the trivial phase ($h=2.0$) and (c-d) the SPT phase ($h=0.5$). The system size is $L=64$ for OBC and $L=128$ for PBC, and $k$ counts the spectrum from the lowest-lying levels. The ground state manifolds of the ES show the same number of levels as the physical ground state manifolds, as indicated by the red boxes.}
\label{fig:gapped}
\end{figure}

As a warm-up, we first study the 1D version of Li-Haldane conjecture in the following cluster SPT model~\cite{verresen2017prb,yu2022prl,guo2022pra}, $H = - \sum_{i=1}^{L} \sigma_{i}^{z}\sigma_{i+1}^{x}\sigma_{i+2}^{z} - h \sum_{i=1}^{L} \sigma_{i}^{x}$. 
Here, the Pauli matrices $\sigma_{i}^{x/z}$ represent the spin-$1/2$ degrees of freedom on site $i$. 
This model hosts a gapped SPT phase at $h \textless h_c$ and a trivial phase at $h \textgreater h_c$~\cite{verresen2017prb}.
While the ground state is unique in the trivial phase, the SPT phase features a four-fold degeneracy from two edges in open boundary condition (OBC). 
Fig.~\ref{fig:gapped}(a) and (c) illustrate the energy spectrum under OBC for the trivial and SPT phases, respectively. 
The degeneracy arises from zero-energy edge states at either end of the chain. 
The ES under periodic boundary conditions (PBC) for the trivial and SPT phases are shown in Fig.~\ref{fig:gapped}(b) and (d), respectively. 
Remarkably, the ``lowest-energy'' structure, highlighted by red boxes in Fig.~\ref{fig:gapped}, of the ES faithfully reproduces the (non) degeneracy of the (trivial) SPT phase, akin to a 1D manifestation of the Li-Haldane conjecture. 
Consequently, the low-lying bulk ES has been widely employed/discussed as a topological fingerprint in the investigation of gapped topological phases of matter~\cite{pollmann2010entanglement,fidkowski2010prl,hsieh2014prl,turner2011prb}.

\emph{ES in symmetry enriched QCPs.}---Until now, different families of gSPTs have been identified in the literature~\cite{li2023decorated,li2023intrinsicallypurely,wen2023classification,wen2023prb}: non-intrinsic gSPTs usually emerge at critical points between SPTs and spontaneous symmetry breaking phases and exhibit a partial set of edge modes from the adjacent gapped SPT; conversely, intrinsically gSPTs are usually stable phases without a gapped counterpart.
For instance, emergent anomalies that protect edge modes in intrinsically  gSPTs could not arise in a gapped phase in the same dimension with the same symmetry.

\begin{figure*}[tb]
    \includegraphics[width=\linewidth]{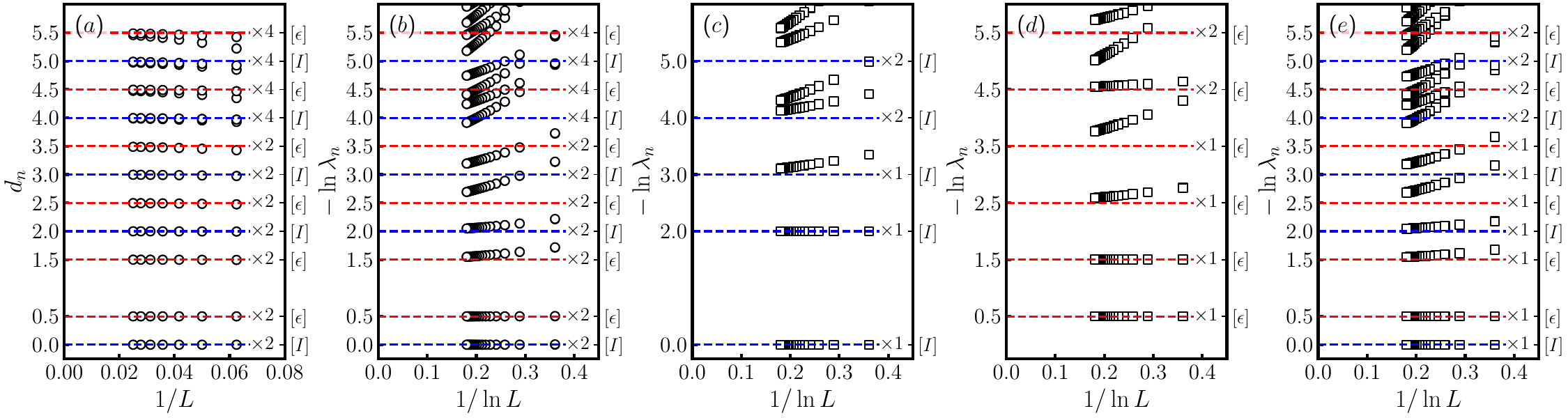}
    \caption{(Color online) (a) OBC energy spectrum and (b) PBC ES of the generalized cluster Ising chain Eq.~\eqref{E1} at the QCP $h=1.0$ for several $L$. 
    The results of the bulk ES with additional projections on the boundary are shown in (c) for $(1+\sigma_{1}^{z})(1+\sigma_{L/2}^{z})$, in (d) for $(1+\sigma_{1}^{z})(1-\sigma_{L/2}^{z})$, and in (e) for $(1+\sigma_{1}^{z})$. All the spectra have been rescaled such that the first two levels are fixed to the corresponding values. Open circles represent a two-fold degeneracy while open squares indicate a single degeneracy.}
    \label{fig:CI}
\end{figure*}

As an example of symmetry enriched QCPs, we consider the 1D generalized cluster Ising model introduced in Ref~\cite{jones2021prr,smith2022prr,verresen2021prx,yu2022prl}, 
\begin{equation}
    \label{E1}
    H_{\text{CI}} = -\sum_{i=1}^{L}\sigma^{z}_{i}\sigma^{x}_{i+1}\sigma^{z}_{i+2}-h\sum_{i=1}^{L}\sigma^{z}_{i}\sigma^{z}_{i+1}\,.
\end{equation}
This model possesses a $\mathbb{Z}_{2}$ spin-flip symmetry and a time-reversal symmetry $\mathbb{Z}^{T}_{2}$: $P = \prod_{i}\sigma^{x}_{i}$ and $T = \mathbb{K}$ (complex conjugation).
By adjusting the tuning parameter $h$, the system undergoes a transition between a ferromagnetic (FM) phase and an SPT phase, with the latter sometimes referred to as the cluster SPT phase. 
The FM-SPT transition is described by a symmetry enriched Ising CFT, where the time-reversal symmetry acts nontrivially on the string operator.
In a semi-infinite geometry, the string operator (symmetry flux) $\sigma_1^z \sigma_2^y \prod_{i=3}^\infty \sigma_i^x$ has a nontrivial charge under time reversal symmetry, making it distinct from a normal Ising CFT. 
Moreover, the charged string operator renders a two-fold degenerate edge mode protected by $\mathbb Z_2 \times \mathbb Z_2^T$ symmetry. 
Intuitively, because $\sigma^z_1$ commutes with the Hamiltonian in a semi-infinite chain, the edge spontaneously breaks the Ising symmetry. 
Note that this is not merely a fine-tuned result but protected by the underlying symmetry~\cite{verresen2021prx}.

To investigate the corresponding bulk-boundary duality, we used the density matrix renormalization group (DMRG) method~\cite{white1992prl,white1993prb,schollwock2005rmp,SCHOLLWOCK201196,cirac2006prb,orus2014ap,stoudenmire2012arcmp} (The details of the algorithm are introduced in the Supplementary Materials (SM)) to calculate the bulk ES and many-body energy spectrum under OBC, respectively, as shown in Fig.~\ref{fig:CI}(a) and (b). 
After a proper rescaling, we observe that i) the ES shows the same doubly degeneracy in OBC energy spectrum, reflecting the nontrivial edge state, and ii) the bulk ES contains the same operator content in the corresponding boundary CFT. 
This example also demonstrates that the bulk wave function nicely encodes the information on the topology and
the operator content under OBC (see Sec. II and III of SM for discussions on other gSPT families).

\begin{figure}[tb]
    \includegraphics[width=0.9\linewidth]{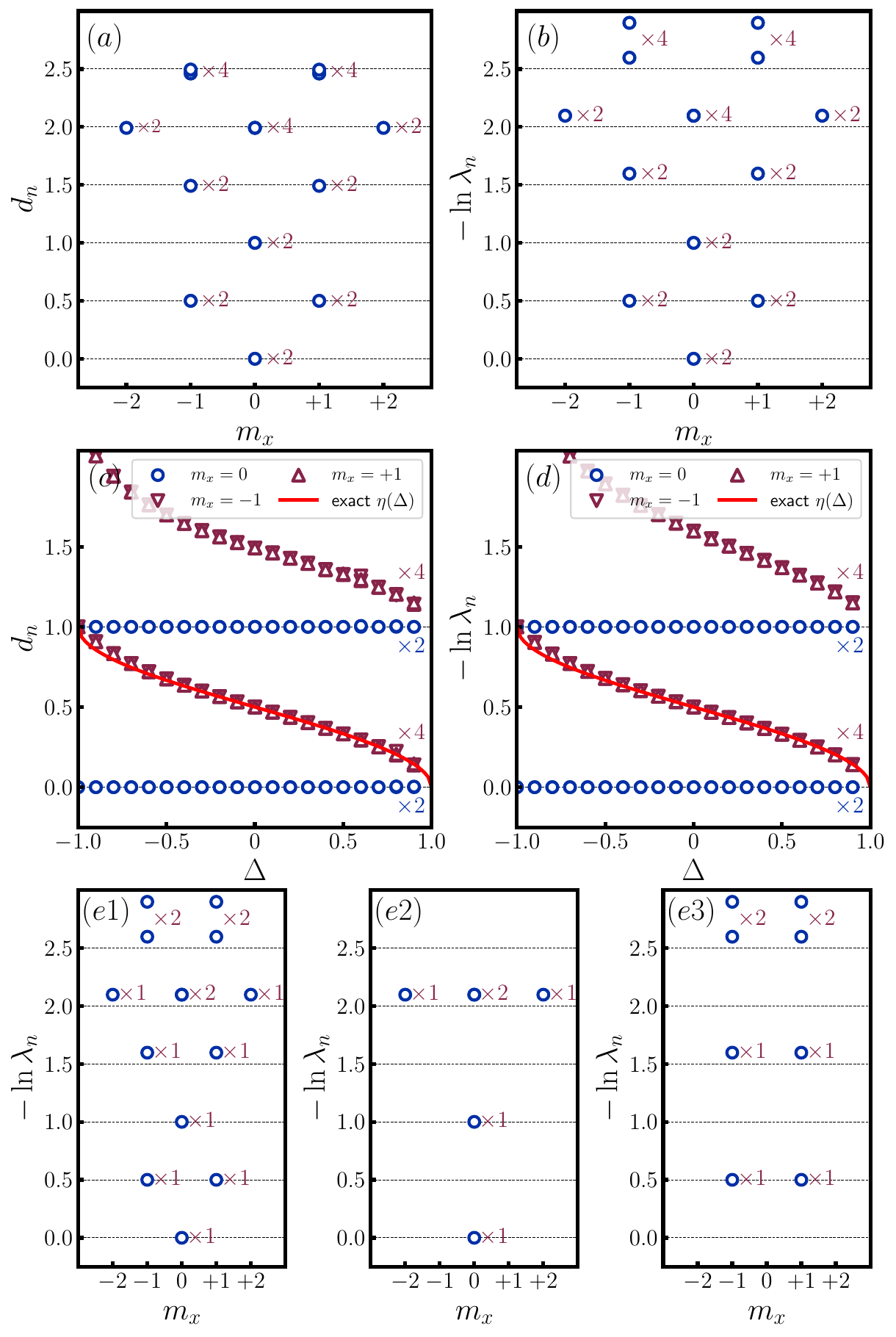}
    \caption{(Color online) (a) OBC energy spectrum and (b) PBC entanglement spectrum labeled by the quantum number $m_{x}$ for the intrinsically  gSPT in the case of $\Delta=0$. 
    The spectrum is rescaled such that the first two levels within the $m_{x}=0$ sector are fixed to $0$ and $1$, respectively. 
    (c-d) The spectrum within the $m_{x}=0$ and $\pm1$ sectors as a function of $\Delta$. 
    The rescaled value of the first level in the $m_{x}=\pm1$ sector is related to the Luttinger parameter and is compared with the exact solution, $\eta(\Delta)=1-\arccos{(-\Delta)}/\pi$ (red solid line). 
    (e1)-(e3) display the resulting entanglement spectrum for $\Delta=0$ after the projection $(1+\sigma_{2L}^{z})$, $(1+\sigma_{L}^{z})(1+\sigma_{2L}^{z})$, and $(1-\sigma_{L}^{z})(1+\sigma_{2L}^{z})$, respectively, from left to right. 
    The simulated system size is $L=24$ for OBC and $L=64$ for PBC; the colored numbers indicate the degeneracy of each level.
    }
    \label{fig:igSPT}
\end{figure}

\emph{ES in intrinsically gapless SPT phases.}---It is natural to inquire whether the universal spectroscopy correspondence can be extended to stable critical phases. 
Now, we examine a representative system of the intrinsically gapless SPT phase given by~\cite{li2023intrinsicallypurely}:
\begin{equation}
    \label{igspt}
    \begin{split}
    H_{\rm igSPT} = - \sum_{i=1}^{L} \Big( \tau_{2i-1}^{z}\sigma_{2i}^{x}\tau_{2i+1}^{z} + \tau_{2i-1}^{y}\sigma_{2i}^{x}\tau_{2i+1}^{y} \\ + \sigma_{2i}^{z}\tau_{2i+1}^{x}\sigma_{2i+2}^{z} + \Delta \tau_{2i-1}^{x}\tau_{2i+1}^{x}  \Big) \, , 
    \end{split}
\end{equation}
where each pair of $(\tau_{2i-1},\sigma_{2i})$ represents the $i$-th unit cell, and the two species of spins per unit cell are represented by Pauli operators $\sigma^{\alpha}$ and $\tau^{\alpha}$. 
$|\Delta|<1$ in the last term denotes the strength of an exactly marginal symmetric perturbation.
This Hamiltonian can be obtained by stacking an Ising-ordered Hamiltonian with an XXZ chain through the Kennedy-Tasaki transformation~\cite{li2023noninvertible,li2023intrinsicallypurely}. 
The low-energy effective theory is described by a $c=1$ free boson CFT. 
The system possesses a $\mathbb{Z}_{4}$ symmetry generated by $U=\prod_{i}\sigma^{x}_{2i}e^{i\frac{\pi}{4}(1-\tau^{x}_{2i-1})}$, which exhibits an emergent anomaly in the low energies. 
Namely, in the low-energy sector, where $\sigma^z_{2i-2} \sigma^z_{2i} = \tau_{2i-1}^x$, the $\mathbb{Z}_{4}$ is approximately $U \sim \prod_{i}\sigma^{x}_{2i}e^{i\frac{\pi}{4}(1-\sigma^z_{2i-2} \sigma^z_{2i})}$, which is the same anomaly on the boundary of a 2+1D Levin-Gu SPT phase~\cite{levin2012prb}. 
This anomaly prevents the system from realizing a unique symmetry-preserving gapped phase. 
Moreover, in an open chain with a length $L$, the square of the low-energy symmetry operator fractionalizes onto each end of the boundary~\cite{wen2023classification,wen2023prb}, $U^2 \sim \tau^x_1 \sigma_2^z \sigma^z_{2L}$. 
This charge locally anticommutes with the $U$ symmetry, protecting a two-fold ground-state degeneracy. 

To investigate the ES of the intrinsically gSPT phase, we first consider $\Delta=0$, where the ground state is an intrinsically gSPT phase, as proven in the literature~\cite{li2023intrinsicallypurely}. 
It is obvious to note that the sublattice magnetization $m_{x} = \frac{1}{2} \sum_{i} \langle\tau_{2i-1}^{x}\rangle$ is a good quantum number for any $\Delta$. 
Consequently, we can categorize the full spectrum into different sectors labeled by $m_{x}$. 
The results of energy and entanglement spectrum are depicted in Fig.~\ref{fig:igSPT}(a) and (b), respectively. 
We observe that the bulk ES not only exhibits the same degeneracy as the OBC energy spectrum but also shares the same structure. 
Both OBC energy spectrum and bulk ES correspond to the operator content of the free boson boundary CFT~\cite{lauchli2013operator}, which suggests that both topological and boundary CFT information can be obtained in the stable critical phase through ES from a bulk wavefunction.

\begin{figure}[tb]
    \includegraphics[width=1.0\linewidth]{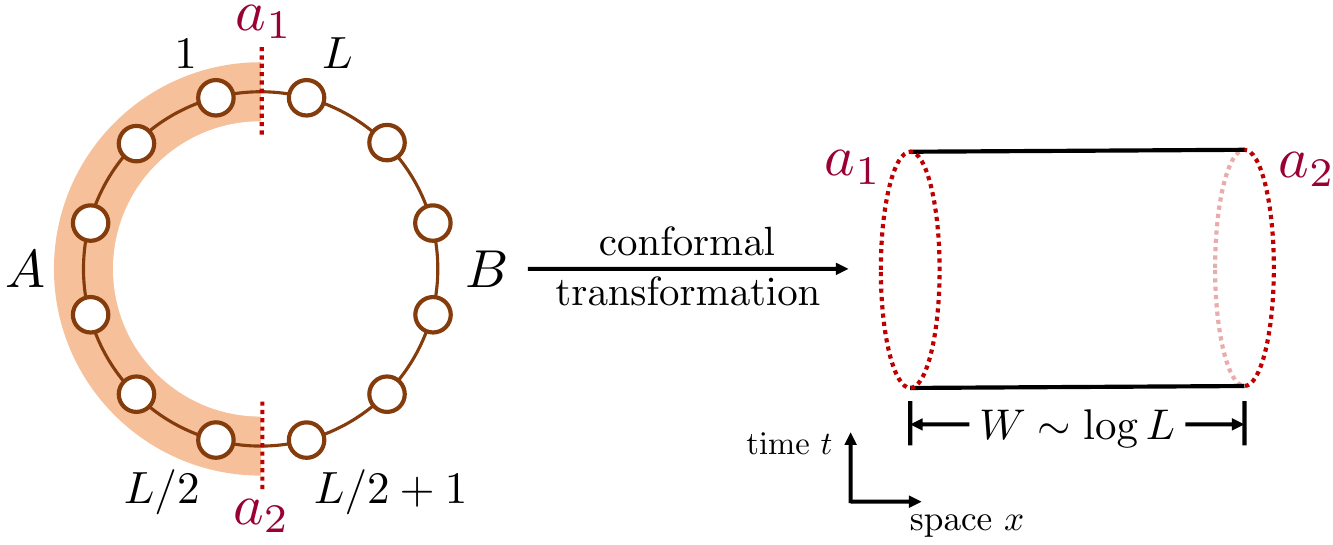}
    \caption{(Color online). The setup involves a bipartition of one-dimensional periodic spin models. 
    The orange shaded region denotes the subsystem $A$, and $B$ represents its complement. 
    The red dotted line represents the entanglement cut, and $a_{1,2}$ labels the boundary condition. 
    After conformal transformation, the reduced density matrix maps to a cylinder (annulus) with width $W \sim \text{log} L$ in the path integral representation~\cite{Ohmori_2015,lauchli2013operator}.}
\label{fig:CFT}
\end{figure}

\emph{ES and boundary CFT.}---There exists an immediate relation between the entanglement Hamiltonian and the Hamiltonian of an open boundary chain~\cite{Cardy_2016,Ohmori_2015}. 
In the continuum limit, the entanglement cut is modeled by a small spatial region of thickness $\epsilon$ at the boundary of $A$ and $B$. 
We consider in our examples the ground state of a one-dimensional periodic chain of a length $L$ with a bipartition $A: (-L/2, L/2)$ and the complement $B$ as shown in Fig.~\ref{fig:CFT}.
The manifold of the Euclidean path integral is given by an infinite cylinder with two entanglement cuts with a radius $\epsilon$. 
A complex number $z = z + L$ presents the cylinder with the imaginary time in the direction of $\text{Im}(z)$, and the entanglement cut at $z= \pm L/2$.
With this topology, the manifold can be mapped onto an annulus that terminates at the entanglement cuts.  
The explicit conformal transformation is given by
\begin{equation}
    \xi(z) = \log \left( \frac{e^{i2\pi z / L} - e^{- i\pi/2}}{e^{i\pi/2 }- e^{i 2\pi z/L}} \right) \, ,
\end{equation}
where $\xi = \xi + 2\pi i$ represents the coordinate of the annulus shown in Fig.~\ref{fig:CFT} with $\xi = x + i t$.  
The two boundaries of the annulus are the conformal image of two entanglement cuts at $\xi \approx \pm \log \frac{2L}{\pi \epsilon} $, which leads to the width of the annulus $W = 2\log \frac{2L}{\pi \epsilon}$.
After this conformal transformation, the entanglement Hamiltonian is then the conformal image that generates the translation in the $\text{Im}(\xi)$ direction. 
Hence, the entanglement spectrum is equivalent to the energy spectrum in the open annulus with the boundary condition given at the entanglement cuts. 
When the ``low-energy" part of the spectrum is concerned, these boundary conditions will flow to conformal boundary condition, which we labeled by $a_1$ and $a_2$. 
Given the boundary states, $a_1$ and $a_2$, and the annulus width $W$, the entanglement spectrum reads 
\begin{equation}
    E^{(a_1, a_2)} = \frac{\pi}{W} \left(- \frac{c}{24} + \Delta_j^{(a_1, a_2)} \right),
\end{equation}
where $\Delta_j^{(a_1, a_2)}$ is the scaling dimension of the allowed operators consistent with the conformal boundary conditions $a_1$ and $a_2$, and $c$ is the central charge of the underlying CFT. Notice that the energy level is inversely proportional to the annulus width, $W$, and via the conformal transformation, the entanglement spectrum is, on the other hand, inversely proportional to $\log L$ (see Sec.IV of SM for a detailed discussion).

Let's elaborate more on the boundary conditions.
A bipartition of Hilbert space is subtle for a quantum field theory, because of the difficulty to associate a Hilbert subspace to a local region~\cite{witten2022does}.
This subtlety can be resolved by considering a finite lattice system with a finite Hilbert space dimension, i.e., there is a finite local Hilbert space at each site $\mathcal H_i$, and then taking the thermodynamic limit by sending the number of sites to infinite. 
With the lattice regularization, different conditions on the entanglement cut can be applied.
For instance, the ``clear-cut" refers to a bipartition of two Hilbert spaces naturally defined by $\mathcal H_A = \otimes_{i\in A} \mathcal H_{i}$, $\mathcal H_B = \otimes_{i\in B} \mathcal H_{i}$ on a lattice system. 
Also, a projection of the wavefunction onto a complete set of commuting operators at the entanglement cut is another way to make the bipartition well-defined in field theory. 
This corresponds to projections on the adjacent site of the entanglement cut in the lattice regularization.

Since the entanglement spectrum can be directly extracted from the ground state from a periodic system, it means the ground state encodes the energy spectrum of an open boundary system.
In the context of gapless symmetry protected topological phase, the open boundary leads to nontrivial edge states located at the boundary.
Therefore, the entanglement spectrum is two-fold degenerate for sufficiently localized edge states. 
This presents the hallmark of the gapless symmetry protected topological phase, making it distinct from the non-topological normal phase.

In the symmetry enriched Ising QCPs of Eq.~\eqref{E1}, the boundary CFT is characterized by a ``superposition'',
$\tilde {\mathbb I} \oplus \tilde \epsilon$, where $\tilde {\mathbb I}$, $\tilde \epsilon$ denotes two fixed boundary conditions in the language of boundary CFT, leading to the operator content $(\tilde {\mathbb I} \oplus \tilde \epsilon) \times (\tilde {\mathbb I} \oplus \tilde \epsilon) = 2 \times ( [\mathbb I] \oplus [\epsilon] ) $ as seen in Fig.~\ref{fig:CI}. 
Here $[\mathbb I], [\epsilon], [\sigma]$ label the operator content of the three primary fields in Ising CFT.
It is in sharp contrast to normal Ising CFT whose boundary state is normally $\tilde \sigma$, i.e., a free boundary condition without double degeneracy.  
The boundary condition beyond the ``clear-cut'' at the entanglement cut provides an additional knob to control the ES. 
We introduce the projection operators at the entanglement cut, $ P_{L,R} \propto (1 \pm \sigma^{z}_{1,L/2})$. 
The projection is applied to the ground state $P_{L,R} \ket{\psi}$, and then the trace over region $B$ will be performed to get entanglement Hamiltonian and spectrum. 
The effect of the projection is to fix the boundary condition to be $\tilde{\mathbb{I}}$ or $\tilde \epsilon$.
As a result, we can modify the ES according to $\tilde{\mathbb{I}} \times \tilde{\mathbb{I}} 
=  [{\mathbb{I}}] $, $\tilde{\mathbb{I}} \times  \tilde \epsilon = [\epsilon] $ and $\tilde{\mathbb{I}} \times (\tilde {\mathbb I} \oplus \tilde \epsilon) = [\mathbb I] \oplus [\epsilon]$ as shown in Fig.~\ref{fig:CI}(c)-(f), respectively. 

{\it Symmetry enriched boundary condition.}---In the intrinsically gapless SPT phases of Eq.~\eqref{igspt}, which is described by a free boson $c=1$ CFT, the boundary condition in an open chain goes beyond the conventional Dirichlet boundary condition~\cite{eggert1992magnetic}. 
Recall the Dirichlet boundary condition in a normal free boson CFT contains states with energy $E_{m_x,n} \sim \frac{1}{W} (\eta(\Delta) m_x^2 + n)$, where $m_x $ ($n$) is an integer labelling the topological sector (the descendant state) and $\eta(\Delta)=1-\arccos{(-\Delta)}/\pi$. 
Here, on the other hand, the boundary state is enriched by the symmetry fractionalization $U^2 = \tau^x_1 \sigma_2^z \sigma^z_{2L}$ 
at both edges, namely, on top of a Dirichlet boundary state, an extra label of the spontaneous magnetization $\sigma_{1,2}$ on each edge needs to be specified.
Hence, the state is enriched, $|m_x , n, \sigma_1, \sigma_2 \rangle $. 
It is worth emphasizing that different from the ``superposition'' of two normal boundary conditions as in the symmetry enriched Ising QCP, here the boundary condition is new and cannot be obtained by a superposition of normal boundary conditions in free boson boundary CFT.
With the presence of an inversion (parity) symmetry of Eq.~\eqref{igspt}, those states are classified by distinct parity: $|2k, n, \sigma, \sigma \rangle $ and $| 2k+1, n, \sigma, -\sigma \rangle $ (see Sec. V of SM for a detailed discussion), each features a double-degeneracy $\sigma = \pm 1$. 
This explains the double degeneracy of ES as seen in Fig.~\ref{fig:igSPT} in the language of boundary CFT.
Going beyond the clear-cut, we can modify the entanglement cut by a projection, i.e., $(1\pm \sigma_{L}^{z})$ or $(1\pm \sigma_{2L}^{z})$.
A projection $(1 + \sigma_{2L}^{z})$ lifts the double degeneracy and results in the remained states: $| 2k, n, 1, 1 \rangle $ and $|2k-1, n, 1, -1 \rangle $;
while a projection $(1+\sigma_{L}^{z})(1+\sigma_{2L}^{z})$ [$(1-\sigma_{L}^{z})(1+\sigma_{2L}^{z})$] allows states only with $m_x \in 2 \mathbb Z$ ($m_x \in 2 \mathbb Z +1$), as clearly shown in Fig.~\ref{fig:igSPT} (e1,2,3), respectively.

\emph{Concluding remarks.}---To summarize, we have investigated several families of 1+1D quantum chains featuring gapless symmetry protected topological states. 
Our primary focus has been to establish a one-to-one correspondence between the bulk ES and the edge energy spectrum, both of which align with the topological degeneracy of the topological state and operator content of the underlying boundary CFT. 
Our finding highlights the universal entanglement spectrum, and thus, opens a new avenue toward understanding of gapless topological phases of matter. 
From the perspective of experimental realization, entanglement spectroscopy holds the potential for implementation in the state-of-the-art quantum simulator platform~\cite{zache2022entanglement,kokail2021prl,joshi2023exploring}.
It has been proposed that the ES can be obtained by learning and subsequently analyzing its spectral properties, which may be useful to demonstrate our findings in experiments. 

\textit{Acknowledgement}: We thank Yijian Zou, Fei Yan, Linhao Li and Da-Chuan Lu for helpful discussions. Numerical simulations were carried out with the ITENSOR package~\cite{itensor} on the Kirin No.2 High Performance Cluster supported by the Institute for Fusion Theory and Simulation (IFTS) at Zhejiang University. X.-J.Yu thank Long Zhang, Limei Xu, Rui-Zhen Huang, Chengxiang Ding and Hong-Hao Song for collaboration on related projects. X.-J.Yu is supported by a start-up grant XRC-23102 of Fuzhou University.
This work is also supported by MOST 2022YFA1402701.
The work of S.-K. J. is supported by a start-up grant from Tulane University.

\bibliography{main}

\clearpage
\onecolumngrid

\newpage
\begin{widetext}

\def\normOrd#1{\mathop{:}\nolimits\!#1\!\mathop{:}\nolimits}
\renewcommand{\(}{\left(}
\renewcommand{\)}{\right)}
\renewcommand{\[}{\left[}
\renewcommand{\]}{\right]}

\def\Eq#1{Eq.~(\ref{#1})}
\def\Fig#1{Fig.~\ref{#1}}
\def\avg#1{\left\langle#1\right\rangle}

\section{Supplemental Material for Universal entanglement spectrum in gapless symmetry protected topological states}

\subsection{Section I: Density matrix renormalization group algorithm and numerical setups}
\label{sec:SM1}

To explore the relationship between the energy spectrum with an open boundary condition (OBC) and the bulk entanglement spectrum from the ground state of a periodic system, we employ the state-of-the-art density matrix renormalization group (DMRG) algorithm~\cite{white1992prl,white1993prb} formulated by matrix product states (MPS)~\cite{cirac2006prb,SCHOLLWOCK201196,orus2014ap}. 
For 1D quantum systems, this variational MPS technique has been proven to be highly reliable and efficient for the study of low-lying physics of strongly correlated many-body systems. 
In the present work, we only focus on 1D quantum lattice models, therefore, DMRG can fulfill our needs quite well.

In addition to the ground state, one of the most important tasks of DMRG, in this study, is to obtain the lowest-lying $k\sim O(10)$ energy levels of a local lattice Hamiltonian $H$. 
To this end, assume that we have computed the lowest-lying $n<k$ eigenstates $\{\ket{\varphi_{i}}\}$ and the associated energy $\{E_{i}\}$.
The eigen-energy is labeled in ascending order, $E_{i} \le E_{i+1}$.
We note that there is at least one eigenstate, namely, the ground state, that we can obtain using the regular DMRG calculation. 
To obtain the $(n+1)$-th energy level, one can add an additional penalty term to the original Hamiltonian
\begin{equation}
    H_{n+1} = H + w \sum_{i=1}^{n} \ket{\varphi_{i}}\!\bra{\varphi_{i}}\,.
\end{equation}
The summation of the projection operators in the second term forms an identity acting on the Hilbert subspace spanned by the first low-lying $n$ eigenstates and raises the energy of these $n$ levels so that the $(n+1)$-th energy level of the original Hamiltonian is now the ground state of the modified Hamiltonian $H_{n+1}$. 
With a sufficiently large value of $w$, a regular DMRG calculation then can be performed to compute the next $(n+1)$-th level by targeting the ground state of $H_{n+1}$~\cite{stoudenmire2012arcmp}. 
Using this strategy, we can, in principle, compute the low-lying excited states one by one to figure out the structure of the open-boundary energy spectrum. 
In our practical simulations, the energy penalty $w$ is set to $50$ and the MPS energy has converged up to the order $10^{-8}$ within DMRG sweeps to achieve high accuracy.

As mentioned above, another task of the simulation is to extract the bulk entanglement spectrum of 1D quantum lattice models. 
In this work, we consider a lattice consisting of $L$ sites (or unit cells each of two sites) with periodic boundary condition (PBC). 
By partitioning the system into two equal parts, namely, $A=\{1,\cdots,L/2\}$ and $B=\{L/2+1,\cdots,L\}$, one can obtain the reduced density matrix for region $A$ by tracing out the degrees of freedom within region $B$,
\begin{equation}
    \label{eq:dm}
    \rho_{A} = {\rm{Tr}}_{B} \ket{\Psi}\!\bra{\Psi} = \sum_{\alpha=1}^{\chi} \lambda_{\alpha} \ket{\phi_\alpha}\!\bra{\phi_\alpha}\,.
\end{equation}
Here, $\ket{\Psi}$ is the many-body ground state of $H$ with PBC and the last equality represents the spectrum decomposition of $\rho_{A}$ with its eigenstates $\ket{\phi_{\alpha}}$ and eigenvalues $\lambda_{\alpha}$. 
Based on this decomposition, the entanglement spectrum is  $\xi_{\alpha}=- \frac1{2\pi} \ln{\lambda_{\alpha}}$ (see~\eqref{eq:entanglement_hamiltonian}). 
In the canonical form of MPS, this information can be extracted efficiently from the singular value decomposition once the ground state $\ket{\Psi}$ has been obtained via regular DMRG calculations. 
Moreover, as presented in the main text, we also consider the change of the entanglement-spectrum structure under the application of local projections on the entanglement cuts (equivalently, the boundary of region $A$), for example, the sites $1$ and $L/2$. 
We note that this operation can also be performed easily in the MPS representation. 
To simulate under PBC, the DMRG calculations were performed with an MPS bond dimension gradually increased up to $\chi_{\rm max}=1024$ and the MPS energy has also converged at least up to the order $10^{-8}$ to guarantee the numerical accuracy.

\subsection{Section II: Entanglement and energy spectrum in a gapless symmetry-protected topological (gSPT) phase}
\label{sec:SM2}

In this and following sections, we aim to check the correspondence between the OBC energy spectrum and PBC entanglement spectrum for several different gapless SPT phases to complement the discussion in the main text. 
Without further indication, here and after, the presented entanglement spectrum of a PBC system is obtained in the ``clear-cut" condition mentioned in Section IV.

The first example is given by the spin-$1$ XXZ model with the Hamiltonian
\begin{equation}
    \label{eq:XXZ}
    H_{\rm XXZ} = \sum_{i=1}^{L} \left( S_{i}^{x}S_{i+1}^{x} + S_{i}^{y}S_{i+1}^{y} + \Delta S_{i}^{z}S_{i+1}^{z} \right)\,,
\end{equation}
where $S_{i}^{\alpha}$ is the $\alpha$ component of the spin-$1$ operator at $i$-th site and $\Delta$ is the anisotropy parameter. 
The phase diagram of this model is very rich, including ferromagnetic (FM) $(\Delta<-1)$, XY $(-1<\Delta<0)$, Haldane $(0<\Delta<\Delta_{\rm c}\approx1.1856)$, and antiferromagnetic (AFM) $(\Delta>\Delta_{\rm c})$ phases by tuning $\Delta$~\cite{chen2003prb}. 
As suggested by Ref.~\cite{verresen2021prx}, the critical point separating the Haldane and AFM phases actually represents a paradigmatic gapless SPT state or symmetry-enriched criticality, with exponentially localized edge modes on the open boundary even when the bulk becomes gapless.

\begin{figure*}[tb]
    \includegraphics[width=0.55\linewidth]{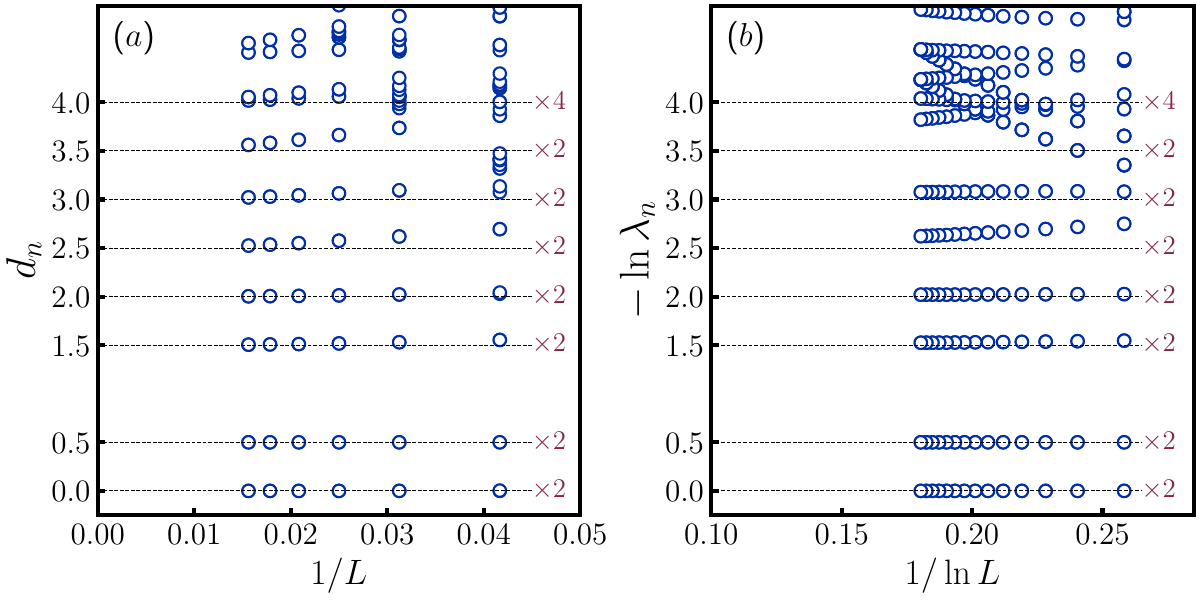}
    \caption{(a) Shifted and rescaled energy spectrum for the spin-$1$ XXZ model at the critical point $\Delta_{\rm c}\approx1.1856$ under OBC for several $L$. The entire spectrum is first shifted so that the first level is fixed to $0$ and then rescaled such that the second level is set to $1/2$. (b) The entanglement spectrum under PBC after the same shifting and rescaling procedure performed in (a). The colored numbers indicate the degeneracy of each level. The deviation of the DMRG data (open circles) from the corresponding dashed lines comes from the finite-size effect and the possible estimation error of the critical point $\Delta_{\rm c}$.}
    \label{fig:XXZ}
\end{figure*}

In Fig.~\ref{fig:XXZ}(a), we present the energy spectrum of the spin-$1$ XXZ model at the critical point $\Delta_{\rm c}$, which belongs to the Ising universality class, under OBC. 
It is interesting to notice that the ground state is doubly degenerate, featuring the non-trivial topological aspect of this criticality. 
To extract more useful information about the criticality and identify the boundary CFT realized by $H_{\rm XXZ}(\Delta_{\rm c})$, the entire energy spectrum is also properly shifted and renormalized to be compared to the operator content of the Ising CFT.
In particular, the first and second levels are set to $0$ and $1/2$, respectively, by defining
\begin{equation}
    \label{eq:rescale}
    d_{n} = \frac{1}{2} \frac{E_{n}-E_{1}}{E_{2}-E_{1}}\,.
\end{equation}
Now, one can clearly see that the energy spectrum shown in Fig.~\ref{fig:XXZ}(a) contains the operator content $2\times (\mathbb{I}\oplus \mathbb{\epsilon})$, due to the presence of gapless edge states.

A theoretical understanding of the observed operator content in the boundary conformal field theory (CFT) language is the following. 
The OBC for this gapless SPT phase is associated with a superposition of boundary conditions $\tilde {\mathbb I} \oplus \tilde \epsilon$ for each of the two boundaries~\cite{verresen2021prx}. 
According to the fusion rule of boundary conditions~\cite{CARDY1986200}, 
\begin{equation}
    \begin{split}
    & \tilde \epsilon \times \tilde \epsilon = \mathbb I, \quad 
    \tilde \epsilon \times \tilde \sigma = \sigma, \quad 
    \tilde \sigma \otimes \tilde\sigma = \mathbb I \oplus \epsilon,
    \end{split}
\end{equation}
where the left-hand side denotes the boundary states and the right-hand side denotes the operator content.
Notice that $\tilde I$ and $\tilde \epsilon$ corresponds to the fixed boundary condition in the Ising model, where the boundary spin is fixed to be up or down, whereas $\tilde \sigma$ corresponds to the free boundary condition. 
It is straightforward to see the operator content $(\tilde {\mathbb I} \oplus \tilde \epsilon) \times (\tilde {\mathbb I} \oplus \tilde \epsilon) = 2 \times ( \mathbb I \oplus \epsilon ) $. 
This is in stark contrast to the normal Ising CFT, where the OBC corresponds to a free boundary condition and the operator content is $\tilde \sigma \otimes \tilde\sigma = \mathbb I \oplus \epsilon$ without a double degeneracy. 

Next, to ascertain that the PBC entanglement spectrum displays the same structure as the OBC energy spectrum, we compute the entanglement spectrum $\xi_{n}$ for the same criticality with an interval length of $L/2$. 
The result after a similar rescaling is shown in Fig.~\ref{fig:XXZ}(b), from which one can indeed find an obvious correspondence between the PBC entanglement and OBC energy spectrum for this case. 
We also observe that the finite-size effect is relatively pronounced in the high-energy part of the entanglement spectrum, primarily attributed to the logarithmic convergence of the entanglement spectrum concerning the system size (see Section IV).

\subsection{Section III: Entanglement and energy spectrum in an intrinsically purely gSPT (ipgSPT) phase}
\label{sec:SM3}

Another example is provided by a quantum spin chain with Pauli operators $\tau$ and $\sigma$ acting, respectively, on odd and even sites, which belongs to the ipgSPT phase by construction. Starting from two decoupled spin-$1/2$ XXZ chains, a Kennedy-Tasaki (KT) transformation can be applied to obtain the corresponding Hamiltonian~\cite{li2023intrinsicallypurely}
\begin{equation}
    \label{eq:ipgSPT}
    H_{\rm ipgSPT} = -\sum_{i=1}^{L} \big(\sigma^{z}_{2i}\tau^{x}_{2i+1}\sigma^{z}_{2i+2}+\sigma^{y}_{2i}\tau^{x}_{2i+1}\sigma^{y}_{2i+2}+h\sigma^{x}_{2i}\sigma^{x}_{2i+2}+ \tau^{z}_{2i-1}\sigma^{x}_{2i}\tau^{z}_{2i+1} + \tau^{y}_{2i-1}\sigma^{x}_{2i}\tau^{y}_{2i+1} + h\tau^{x}_{2i-1}\tau^{x}_{2i+1}\big)\,.
\end{equation}
It has been shown that the ipgSPT corresponds to the parameter region $0<h<1$ and we choose $h=1/2$ here for simulation. As far as we know, this is the first and the only example of ipgSPT phase in the existing literature. The main goal of this section is to verify the correspondence between the entanglement and energy spectrum in this case; for a detailed discussion of the non-trivial topological features of $H_{\rm ipgSPT}$, one can refer to the original paper~\cite{li2023intrinsicallypurely}.

\begin{figure*}[tb]
    \includegraphics[width=0.55\linewidth]{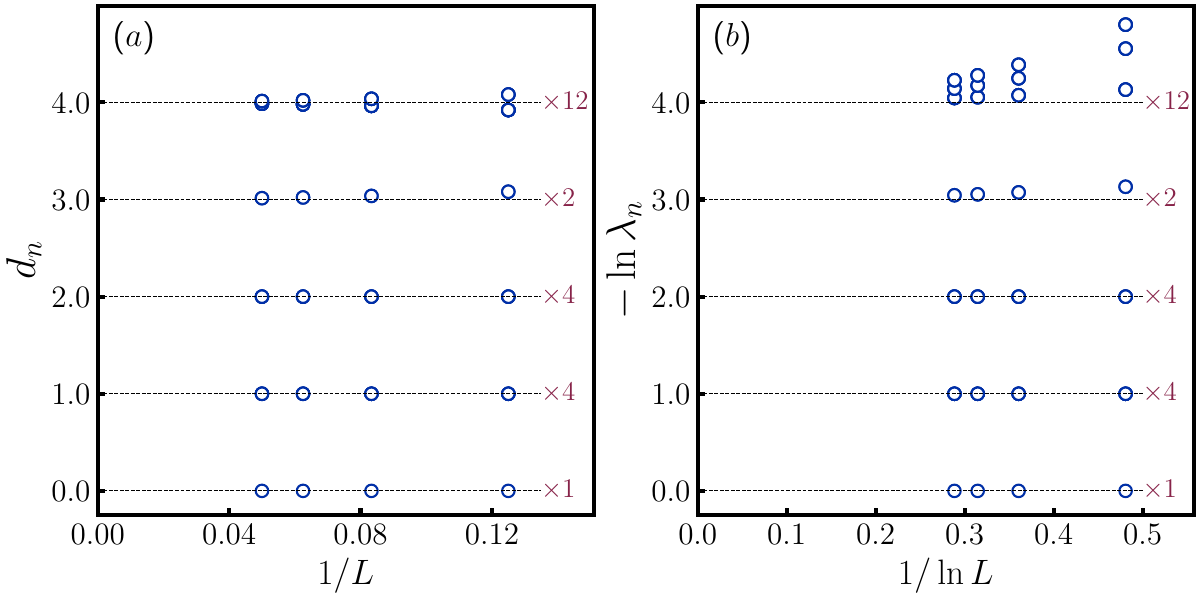}
    \caption{(a) OBC energy spectrum of the ipgSPT with $h=1/2$ for different chain length. The entire spectrum is shifted and rescaled such that the first level is fixed to $0$ and the second level is set to $1$. (b) The entanglement spectrum under PBC after the same rescaling process. The colored numbers indicate the degeneracy of each level. The deviation of the DMRG data (open circles) from the corresponding dashed lines results from the finite-size effect.}
    \label{fig:ipgSPT}
\end{figure*}

In Fig.~\ref{fig:ipgSPT}, we calculate the OBC energy and PBC entanglement spectrum of $H_{\rm ipgSPT}$ at $h=1/2$ for several system sizes $L$. To have a direct comparison, the results have been shifted and rescaled such that the first and second levels are fixed to $0$ and $1$, respectively. It is obvious that the energy and entanglement spectrum share the same structure with each other at least for the displayed low-lying levels and there exists a one-to-one correspondence between them. 

However, it is still not clear how to identify the boundary CFT operator content that the spectrum shown in Fig.~\ref{fig:ipgSPT} corresponds to. The reason can be two-fold. First, as the starting point of $H_{\rm ipgSPT}$ before the Kennedy-Tasaki (KT) transformation is two decoupled spin-$1/2$ XXZ chains, the structure of the spectrum should reflect the coexistence of these two boundary CFTs and can be very complicated. Second, it is suggested that the edge spectrum of $H_{\rm ipgSPT}$ scales as $O(1/L)$ which is comparable with the bulk finite-size gap $\sim1/L$ even in the thermodynamic limit~\cite{li2023intrinsicallypurely}. Therefore, the mixture of the edge and bulk levels means that there is no stable edge mode and makes the understanding of the entire spectrum even more difficult. Although it is interesting to understand the boundary CFT realized by the spectrum, this task is not the main focus of our work and we leave it for future studies.

\subsection{Section IV: Entanglement Hamiltonian in conformal field theory}
In this section, we briefly review the relation between the entanglement Hamiltonian and the Hamiltonian of an open boundary chain~\cite{Cardy_2016, Ohmori_2015}.
Given a bipartition of the Hilbert space $\mathcal H = \mathcal H_A \otimes \mathcal H_B$, the entanglement Hamiltonian of the subsystem $A$ of a density matrix is defined by
\begin{equation} \label{eq:entanglement_hamiltonian}
    K_A = - \frac1{2\pi} \log \rho_A,
\end{equation}
where $\rho_A = \Tr_{\mathcal H_B} \rho$ is the reduced density matrix of the subsystem $A$. 
Note that the inclusion of a $(2\pi)^{-1}$ factor is only for convenience. 
A bipartition of Hilbert space is subtle for a quantum field theory, because of the difficulty to associate a Hilbert subspace to a local region~\cite{witten2022does}. 
This subtlety can be resolved by considering a finite lattice system with a finite Hilbert space dimension, i.e., there is a finite local Hilbert space at each site $\mathcal H_i$, and then taking the thermodynamic limit by sending the number of sites to infinite. 
Further, different conditions on the entanglement cut can be applied.
For instance, if the cut between the two Hilbert spaces (entanglement cut) is at a bond, the bipartition of two Hilbert spaces is naturally defined by 
\begin{equation}
    \mathcal H_A = \otimes_{i\in A} \mathcal H_{i}, \quad \mathcal H_B = \otimes_{i\in B} \mathcal H_{i}.
\end{equation}
This type of cut is referred to as ``clear-cut". 
If the entanglement cut is at a site $a$, then a projection onto the complete set of commuting operator at that site can be applied to the density matrix
\begin{equation}
    \rho' = P_a \rho P_a,
\end{equation}
where $P_a$ denotes the corresponding projection.

In the continuum limit, the entanglement cut is modeled by a small spatial region of thickness $\epsilon$ at the boundary of $A$ and $B$. 
To be concrete, we consider the ground state of a one dimensional (1D) conformal field theory with a length $L$ and a PBC. 
The bipartition is that $A: (-L_A/2, L_A/2)$, and $B$ is the complement.  
The manifold of the Euclidean path integral is given by an infinite cylinder with two entanglement cuts with a radius $\epsilon$.
With this topology, the manifold can be mapped onto an annulus that terminates at the entanglement cuts as shown in the following,
\begin{figure}
    \centering
    \includegraphics[width=0.5\textwidth]{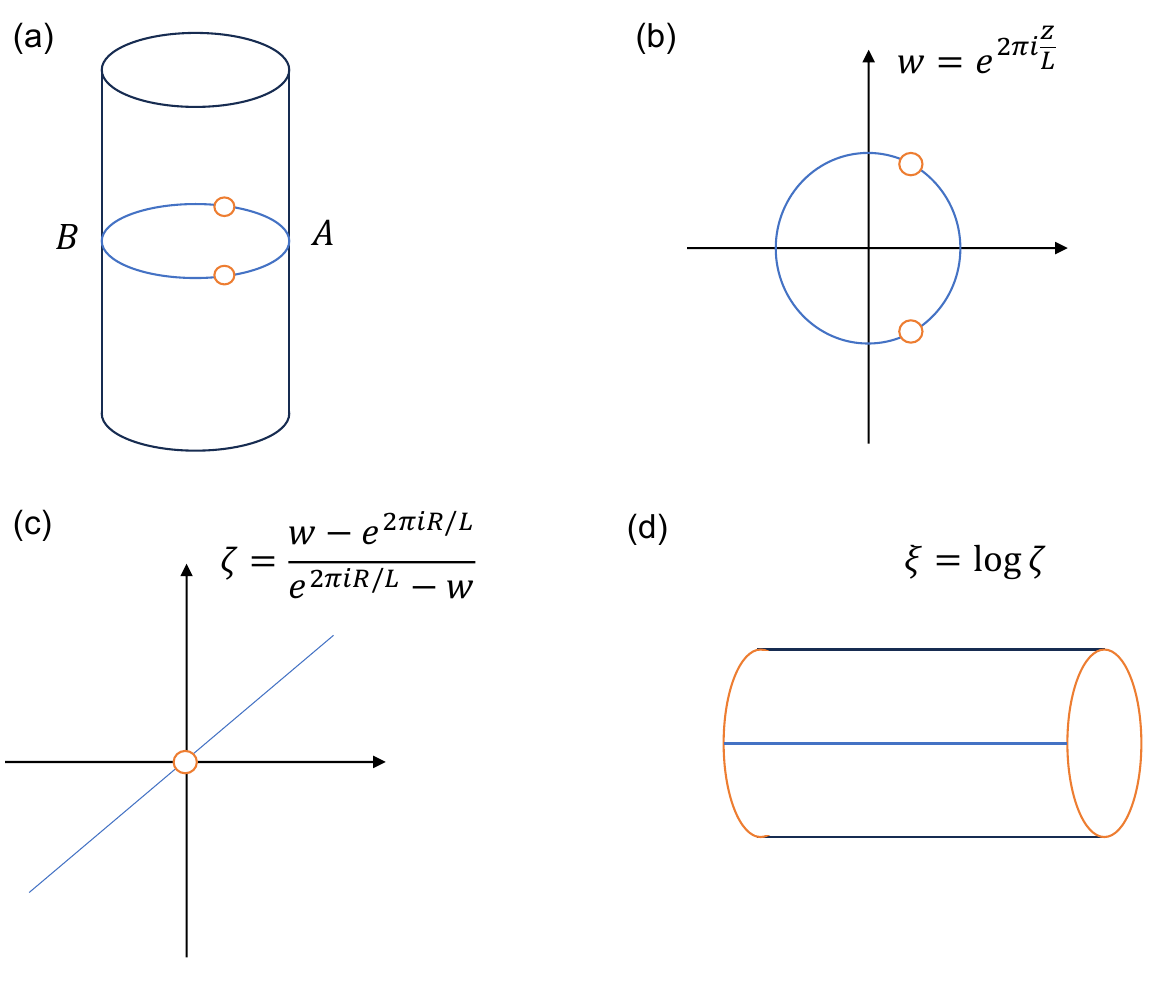}
\end{figure}
Here the cylinder (a) is parameterized by a complex number $z = z + L$. 
The blue circle (orange holes) represents the time reversal invariant slice (entanglement cuts). 
The position of entanglement cut is $z= \pm L_A/2$. 
A series of conformal mappings indicated in the figure will send the cylinder with two cuts to an annulus. 
The explicit conformal transformation is given by
\begin{equation}
    \xi(z) = \log \left( \frac{e^{2\pi i z / L} - e^{- \pi i L_A/L}}{e^{\pi i L_A/L }- e^{2\pi i z/L}} \right)
\end{equation}
with $\xi = \xi + 2\pi i$. 
The two boundaries of the annulus are the conformal image of two entanglement cuts at $\xi \approx \pm \log \frac{2L\sin \pi L_A/L}{\pi \epsilon} $, which leads to the width of the annulus $W = 2\log \frac{2L\sin \pi L_A/L}{\pi \epsilon}$.
After this conformal transformation, the entanglement Hamiltonian $K_A$ is then the conformal image that generates the translation in the $\text{Im}[\xi]$ direction. 
Hence, the entanglement spectrum is equivalent to the Hamiltonian with the boundary condition given at the entanglement cuts.
These boundary conditions will flow to conformal boundary states, which we denote by $a$ and $b$. 
Given the boundary states, $a$ and $b$, and the annulus width $W$, the entanglement spectrum reads 
\begin{equation}
    E^{(ab)} = \frac{\pi}{W} \left(- \frac{c}{24} + \Delta_j^{(ab)} \right),
\end{equation}
where $\Delta_j^{(ab)}$ is the scaling dimension of the allowed operators consistent with the conformal boundary conditions $a$ and $b$. 
Notice that the energy level is inversely proportional to the annulus width, $W$, and via the conformal transformation, the entanglement spectrum is, on the other hand, inversely proportional to $\log L_A$. 

The entanglement spectrum can be directly extracted from the ground state from a periodic system. 
So, such a ground state wave function encodes the energy spectrum of an open boundary system.
In the context of gapless symmetry protected topological phase, the open boundary leads to nontrivial edge states located at the boundary.
Therefore, the entanglement spectrum is two-fold degenerate for sufficiently localized edge states.  
This presents the hallmark of the gapless symmetry protected topological phase, making it distinct from the non-topological normal phase.

\subsection{Section V: Additional results for the intrinsically gSPT (igSPT) phase}
\label{sec:SM4}

In this section, we provide additional results for the igSPT phase discussed in the main text, which is a spin chain with two spin-$1/2$, namely, $\tau_{2i-1}$ and $\sigma_{2i}$, per unit cell described by the following Hamiltonian~\cite{li2023decorated,li2023intrinsicallypurely}
\begin{equation}
    \label{eq:igSPT}
    H_{\rm igSPT} = - \sum_{i=1}^{L} \big( \tau_{2i-1}^{z}\sigma_{2i}^{x}\tau_{2i+1}^{z} +  \tau_{2i-1}^{y}\sigma_{2i}^{x}\tau_{2i+1}^{y} + \sigma_{2i}^{z}\tau_{2i+1}^{x}\sigma_{2i+2}^{z} + \Delta \tau_{2i-1}^{x}\tau_{2i+1}^{x} \big)\,,
\end{equation}
where we have included a perturbation term, $\sum_{i} \Delta \tau_{2i-1}^{x}\tau_{2i+1}^{x}$.

\begin{figure*}[tb]
    \includegraphics[width=0.75\linewidth]{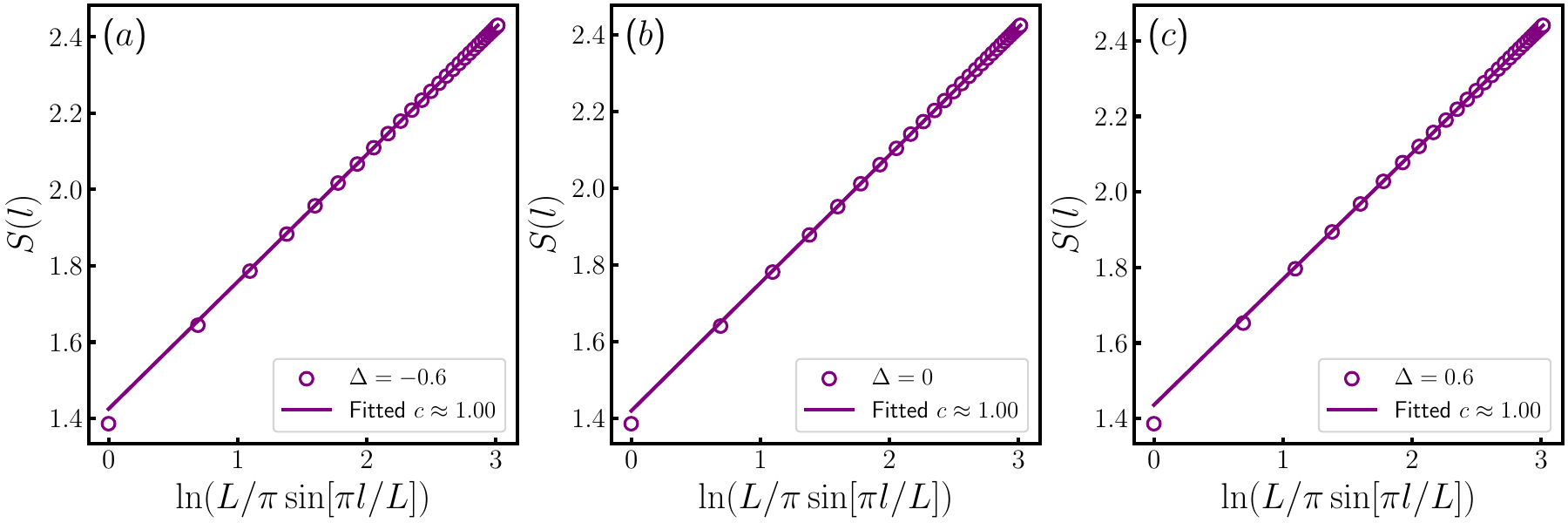}
    \caption{Entanglement entropy of an interval of length $l$ is fitted by the celebrated formula~\cite{Cardy2004charge,Cardy2009charge}: $S(l) = c/3 \ln(L/\pi \sin[\pi l/L]) + c'$, where $c$ is the central charge of the underlying CFT and $c'$ is a non-universal quantity. The system size used in the simulation is $L=64$ and PBC is adopted. It is obvious that $c=1$ for $\Delta=-0.6$, $0$, and $0.6$, collaborating that the low-energy theory for $H_{\rm igSPT}$ is a free boson CFT.}
    \label{fig:igSPT_C}
\end{figure*}

\begin{figure*}[tb]
    \includegraphics[width=0.55\linewidth]{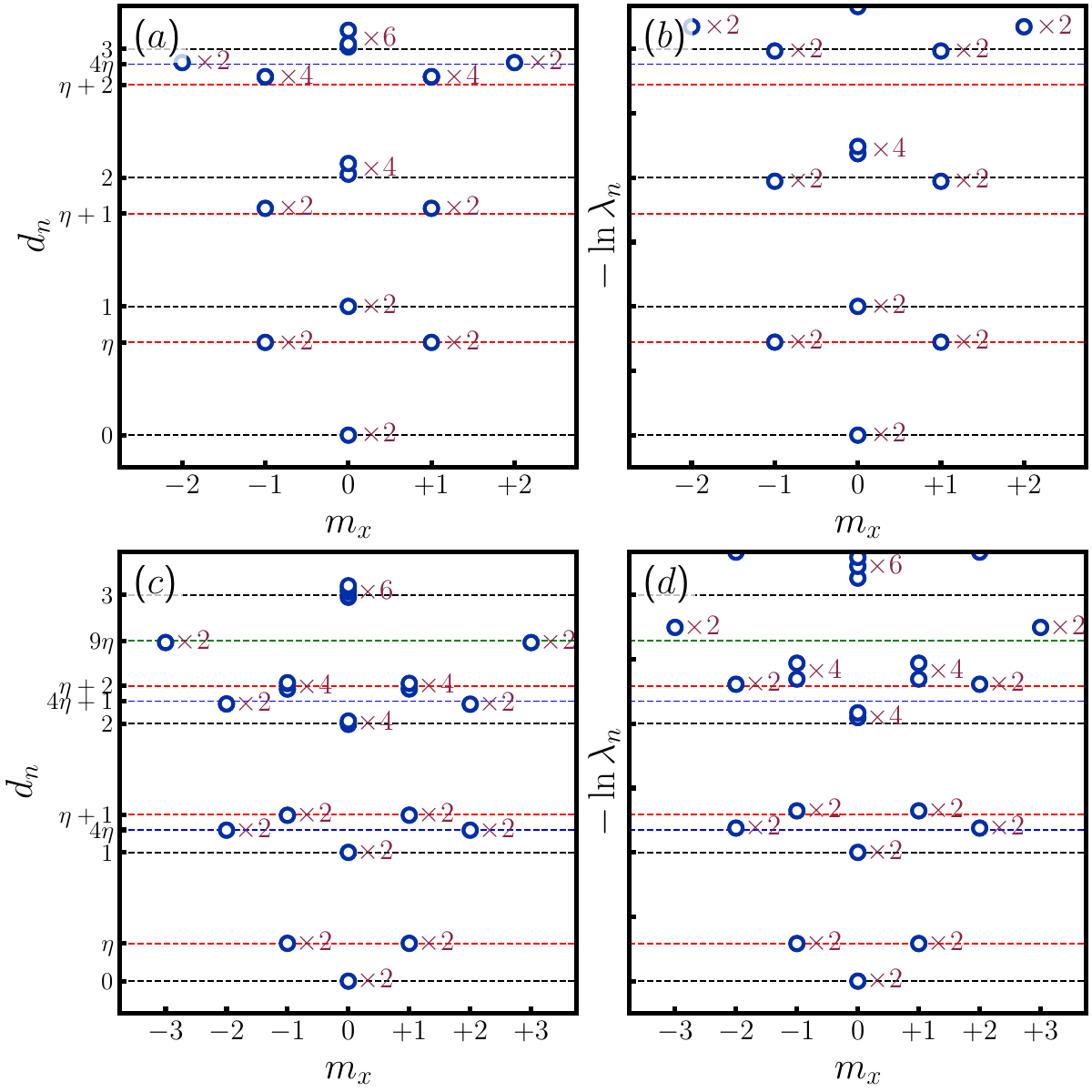}
    \caption{OBC energy and PBC entanglement spectrum of $H_{\rm igSPT}$ for (a)-(b) $\Delta=-0.6$ and (c)-(d) $\Delta=0.6$\,. The entire spectrum has been properly rescaled such that the first and second levels within the $m_{x}=0$ sector are set to $0$ and $1$, respectively. Based on this procedure, the rescaled value of the first level in the $m_{x}=\pm1$ sector gives the scaling dimension of the first non-identity primary field, $\eta$, which equals to $1/2$ for $\Delta=0$ (see the results in the main text), $0.720$ for $\Delta=-0.6$, and $0.294$ for $\Delta=0.6$\,. The deviation of the relatively ``high-lying'' spectrum from the corresponding dashed lines is caused by the finite-size effect. Simulated system sizes are $L=24$ for OBC and $L=64$ for PBC.}
    \label{fig:igSPT_EE}
\end{figure*}

\begin{figure*}[tb]
    \includegraphics[width=0.85\linewidth]{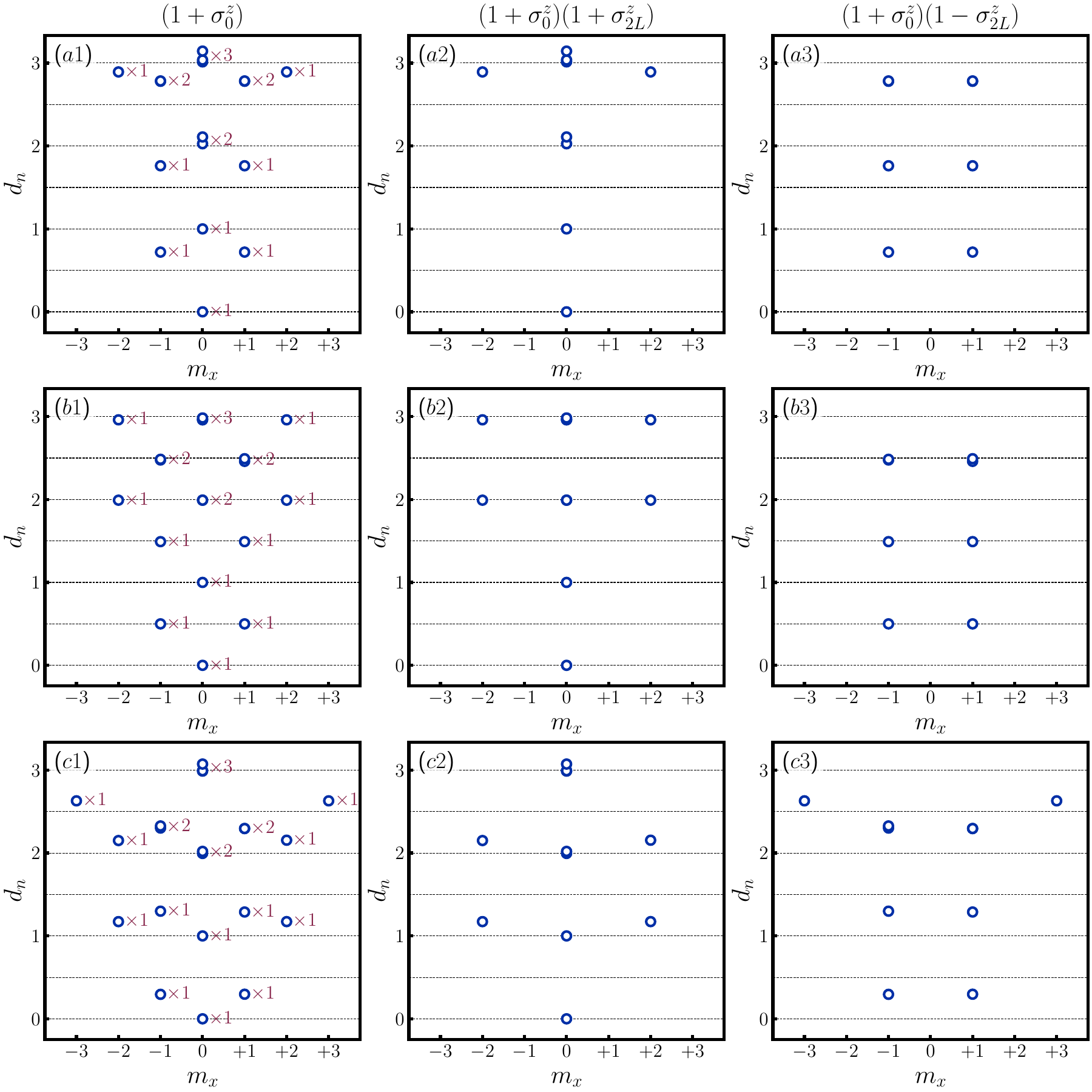}
    \caption{Change of the OBC energy spectrum with an additional projection, specifically, $(1+\sigma_{0}^{z})$ for the first column, $(1+\sigma_{0}^{z})(1+\sigma_{2L}^{z})$ for the second column, and $(1+\sigma_{0}^{z})(1-\sigma_{2L}^{z})$ for the third column. The colored numbers indicate the degeneracy of the corresponding level. Parameters: $\Delta=-0.6$ for (a1)-(a3), $\Delta=0.0$ for (b1)-(b3), and $\Delta=0.6$ for (c1)-(c3). The simulation is performed under OBC with system size $L=24$  directly for Eq.~\eqref{eq:igSPT_open}.}
    \label{fig:igSPT_Proj_Energy}
\end{figure*}

\begin{figure*}[tb]
    \includegraphics[width=0.85\linewidth]{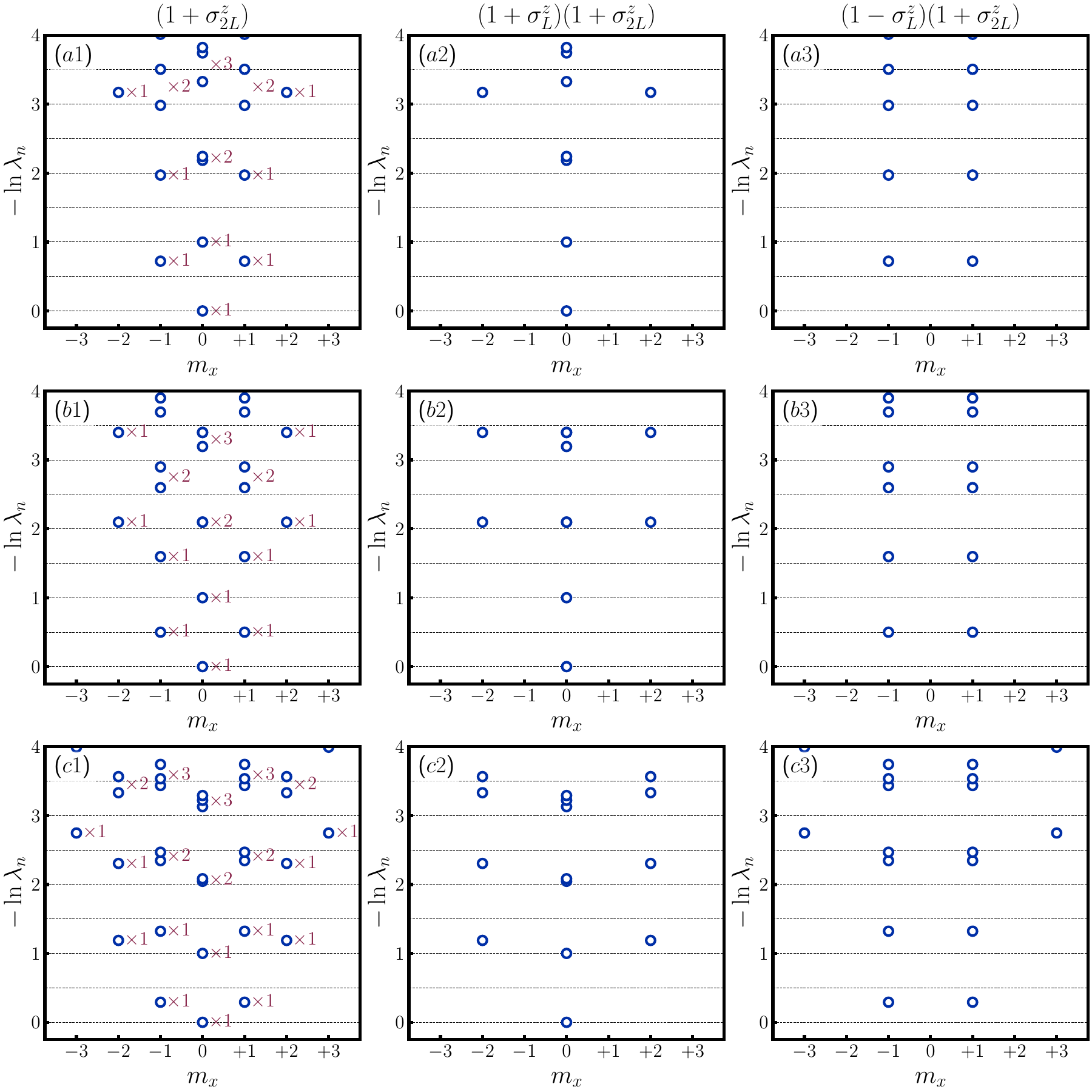}
    \caption{Change of the bulk entanglement spectrum with an additional projection, specifically, $(1+\sigma_{2L}^{z})$ for the first column, $(1+\sigma_{L}^{z})(1+\sigma_{2L}^{z})$ for the second column, and $(1-\sigma_{L}^{z})(1+\sigma_{2L}^{z})$ for the third column. The colored numbers indicate the degeneracy of the corresponding level. Parameters: $\Delta=-0.6$ for (a1)-(a3), $\Delta=0.0$ for (b1)-(b3), and $\Delta=0.6$ for (c1)-(c3). System size simulated is $L=64$ under PBC.}
    \label{fig:igSPT_Proj_Entanglement}
\end{figure*}

Recall that the igSPT model before the KT transformation is an Ising chain (without transverse field) stacked with a spin-$1/2$ XXZ chain,
\begin{equation}
    H_\text{XXZ} = - \sum_{i=1}^L \left( \tau_{2i-1}^y \tau_{2i+1}^y + \tau_{2i-1}^z \tau_{2i+1}^z + \Delta \tau_{2i-1}^x \tau_{2i+1}^x \right) \, .
\end{equation}
It is well known that the XXZ model is exactly solvable.
In bosonization language, it is described by a free boson theory with the Lagrangian density
\begin{equation}
    \mathcal L = \frac12 \left( \frac1{v_F} (\partial_\tau \phi)^2 + v_F (\partial_x \phi)^2 \right),
\end{equation}
where $\phi$ denotes the boson field and $\partial_\mu = \partial_{x}, \partial_\tau$. 
The spin operators are expressed by
\begin{equation} \label{eq:spin_bosonization}
    \begin{split}
        \frac12 \tau_i^x &\approx - \frac1{2\pi R} \partial_x \phi + A (-1)^i \sin \frac\phi{R}, \\
        \frac12 \tau_i^y &\approx \cos (2\pi R \theta) \left( C + B (-1)^i \cos \frac{\phi}R \right), \\
        \frac12 \tau_i^z &\approx \sin (2\pi R \theta) \left( C + B (-1)^i \cos \frac{\phi}R \right),
    \end{split}
\end{equation}
where $\phi$ ($\theta$) denotes the free (dual) boson and $A, B, C$ are non-universal constants.
The radius $R$ of the bosonic field is determined by the anisotropy parameter $\Delta$, $R= \sqrt{\frac1{2\pi} - \frac1{2\pi^2} \arccos (-\Delta)}$.

The boundary condition of the XXZ model with an open boundary corresponds to the Dirichlet boundary condition of the field $\phi$, giving rise to the partition function~\cite{eggert1992magnetic}
\begin{equation}
    Z_\text{XXZ} = \Tr[e^{-\beta H_\text{XXZ}}] = \frac1{\eta(q)} \sum_{m_x \in \mathbb Z} q^{ 2\pi R^2 m_x^2} =\sum_{m_x \in \mathbb Z} \sum_{n =1}^\infty p(n) e^{2\pi R^2 m_x^2 + n}
\end{equation}
where $q = e^{-\beta \pi v_F/L}$ and $\eta(q) = q^{1/24} \prod_{n=1}^\infty (1-q^n)$ is the Dedekind $\eta$-function.
$p(n)$ is the number of partitions of the integer $n$, e.g., $p(1) = 1, p(2) = 1, p(3) = 2, p(4) = 3, p(5) = 5 ,...$\,.
From the partition function, the operator content is classified into different topological sectors given by an integer $m_x$, and within each topological sector, another integer $n$ denotes the excited state.
Hence, in the XXZ sector, the eigenstate $|m_x, n \rangle$ is labeled by two integers $m_x$ and $n$ with
the eigenenergy 
\begin{equation}
    E_{m_x,n} = \frac{\pi v_F}{L} (2\pi R^2 m_x^2 + n) .
\end{equation}
and the degeneracy $p(n)$. 
The topological number $m_x$ is associated with the total spin $m_x = \sum_i \frac12 \tau_{2i-1}^x$, which is a good quantum number in the XXZ model.
If the number of sites is even (odd), $m_x$ is an integer (half-integer). 
Before the KT transformation because the XXZ model and the Ising ferromagnet are decoupled, there will be a trivial doubling for all states due to the two possible magnetizations.

Because the KT transformation is unitary with an open boundary condition, after the KT transformation, the eigenvalue will not change. 
One can expect that its OBC energy spectrum contains the operator content of a free boson CFT with $c=1$ central charge (see Fig.~\ref{fig:igSPT_C}). 
Therefore, the operator content is still doubly degenerate. 
However, this degeneracy is nontrivial because now the $\sigma$ spins and the $\tau$ spins are strongly coupled in~\eqref{eq:igSPT}. 
Before we discuss this novel degeneracy, let's set the background and introduce the relevant symmetry for later convenience. 
For clarity, we consider a chain with $2L+1$ sites, $i=0,...,2L$, where the $\sigma$ ($\tau$) spin is located at even (odd) site. 
The model has an open boundary condition that terminates on the $\sigma$ spins at the site $0$ and the site $2L$, 
\begin{equation}
    \label{eq:igSPT_open}
    H_{\rm igSPT} = - \sigma_0^z \tau_1^x \sigma_2^z - \sum_{i=1}^{L-1} \big( \tau_{2i-1}^{z}\sigma_{2i}^{x}\tau_{2i+1}^{z} +  \tau_{2i-1}^{y}\sigma_{2i}^{x}\tau_{2i+1}^{y} + \sigma_{2i}^{z}\tau_{2i+1}^{x}\sigma_{2i+2}^{z} + \Delta \tau_{2i-1}^{x}\tau_{2i+1}^{x} \big)  .
\end{equation}
We also consider $L$ to be an even number, $L \in 2 \mathbb Z$. 
The model respects a parity symmetry,
\begin{equation}
    I: \quad \tau^\alpha_{2i-1} \rightarrow \tau^\alpha_{2L-(2i-1)}, \quad \sigma^\alpha_{2i} \rightarrow \sigma^\alpha_{2L-2i}.
\end{equation}
Because $L$ is an even integer, the middle site $i=L$ is the $\sigma$ spin, the parity is a site (bond) parity for the $\sigma$ ($\tau$) spin.
The presence of bond parity symmetry classifies the eigenstate into even and odd parity sectors. 
The parity is closely related to the topological sector $m_x$~\cite{eggert1992magnetic}, 
\begin{equation}
    I = (-1)^{m_x}.
\end{equation}

There is an easy way to see such a degeneracy in the model~\eqref{eq:igSPT}.
The boundary $\sigma$ spins, $\sigma^z_{0}$ and $\sigma^z_{2L}$, commute with the Hamiltonian~\eqref{eq:igSPT_open}. 
These two boundary spins are related via the parity transformation.
Therefore, the eigenstates can be labeled by $|m_x, n, \sigma_0, \sigma_{2L} \rangle$, which is further classified into two sectors, i.e., even parity:
$|m_x = 2k, n, \sigma, \sigma \rangle $ and, odd parity, $|m_x = 2k+1, n, \sigma, \bar \sigma \rangle $.
Here $\sigma = -\bar \sigma = \pm 1$.
The boundary spins $\sigma^z_0$ and $\sigma^z_{2L}$ anticommute with the $\mathbb Z_4$ symmetry introduced in~\cite{li2023decorated}, $U=\prod_{i}\sigma^{x}_{2i}e^{i\frac{\pi}{4}(1-\tau^{x}_{2i+1})}$, so every eigenstate is two-fold degenerate.
Namely, $|m_x = 2k, n, \sigma, \sigma \rangle $ for $\sigma = \pm 1$ is degenerate and $|m_x = 2k, n, \sigma, \bar\sigma \rangle $ is also degenerate for $\sigma = \pm 1$.
In summary, in the presence of inversion symmetry, the spectrum of an open chain given by~\eqref{eq:igSPT_open} can be classified into even and odd parity with energy $E_{m_x=2k, n}$ and $E_{m_x=2k+1, n}$, respectively, and all levels are doubly degenerate. 
It is worth pointing out that the degeneracy is a consequence of the gapless edge state protected by the $\mathbb{Z}_4$ symmetry, irrespective of the parity symmetry. 
The presence of the parity symmetry relates the degenerate state with $\sigma_{0} = \sigma_{2L}$ ($\sigma_{0} = - \sigma_{2L}$) to topological sector $m_x=2k$ ($m_x = 2k+1$).

With this theoretical understanding, the projection of the boundary spin $(1\pm \sigma_0^z)$ and $(1 \pm \sigma_{2L}^z)$ can be easily understood. 
For a single projection on one edge, $1+\sigma_0^z$, the degeneracy is lifted, because only a single non-degenerate edge state is allowed. 
The parity sector is determined solely by the degree of freedom on the other edge, $\sigma_{2L}^z$. 
A joint projection on both edges, $(1+\sigma_0^z)(1+\sigma_{2L}^z) $, selects even parity states, $w \in 2 \mathbb Z$.
While the other one, $(1+\sigma_0^z)(1-\sigma_{2L}^z) $, selects odd parity states, $w \in 2 \mathbb Z + 1$.
This explains the operator content seen in Fig.~\ref{fig:igSPT_Proj_Energy}, where we have performed a direct simulation of Eq.~\eqref{eq:igSPT_open} with boundary projections applied on the energy spectrum to achieve the parity selection described here.

Having understood the doubly degeneracy in model~\eqref{eq:igSPT} by including another $\sigma$ spin at the site 0, now we come back to a normal open boundary condition.
In Figs.~\ref{fig:igSPT_EE}(a) and (c), we present the OBC energy spectrum of the igSPT phase for $\Delta=-0.6$ and $0.6$, respectively. 
After a standard rescaling procedure used in Ref.~\cite{lauchli2013operator}, which sets the first and second levels in the sector $m_{x}=0$, respectively, to $0$ and $1$, we can indeed see that the lowest-lying levels in each sector form a parabolic envelope; there is a doubly degenerate level with rescaled value $m_x^2\eta$ for each sector $m_{x}$. 
Here, $\eta$ is the scaling dimension of the first non-identity primary field (labeled by $m_{x}=\pm1$) estimated by
\begin{equation}
    \eta = \frac{E_{1}(\pm1)-E_{1}(0)}{E_{2}(0)-E_{1}(0)}\,,
\end{equation}
where $E_{k}(m)$ is the $k$-th energy level in the $m_{x}=m$ sector. The value of $\eta$ is related to the Luttinger liquid parameter and its dependence on the anisotropy $\Delta$ is known exactly: $\eta = 1 - \arccos{(-\Delta)}/\pi$. It is noted that the degeneracy of all levels are doubled compared with the spectrum of a pure spin-$1/2$ XXZ chain (see the schematic representation Fig.2(d) in Ref.~\cite{lauchli2013operator}) highlighting the non-trivial edge modes living on the boundary.

To see if the bulk entanglement spectrum can reflect the same information shown above, we calculate the PBC entanglement spectrum, whose rescaled result is displayed in Figs.~\ref{fig:igSPT_EE}(b) and (d) for $\Delta=-0.6$ and 0.6, respectively. Here, the spectrum $\{\xi_{k}\}$ is labeled by the corresponding magnetization within the region $A$ [see Eq.~\eqref{eq:dm}],

\begin{equation}
    m^{k}_{x} = \frac{1}{2} \sum_{i=1}^{L/2} \bra{\phi_{k}} \tau_{2i-1}^{x} \ket{\phi_{k}}\,.
\end{equation}

Now, we can see that the bulk entanglement spectrum recovers the same structure as the OBC energy spectrum. The deviation of the relatively ``high-lying'' part of the spectrum is due to the possible finite-size effect.

In addition, we also investigate the response of the entanglement spectrum to the application of projections. Here, the projections are acted on $\sigma_{L}$ and $\sigma_{2L}$, with the specific form 
\begin{equation}
P = (1\pm\sigma_{L}^{z})(1\pm\sigma_{2L}^{z})\,.
\end{equation}
As shown in Fig.~\ref{fig:igSPT_Proj_Entanglement}, it is found that a single projection on $\sigma_{L}$ or $\sigma_{2L}$ can remove the doubly degeneracy but the resulting spectrum can still reflect the same operator content of the boundary CFT (see Fig.~3 in the main text and Fig.~\ref{fig:igSPT_EE}). Moreover, when the projections on $\sigma_{L}$ and $\sigma_{2L}$ are simultaneously applied, depending on the spin orientations chosen for sites $L$ and $2L$, the contribution from odd or even quantum sectors can be further projected out from the degeneracy-removed entanglement spectrum. The results reveal how implementations on the entanglement cut can effect the structure of the bulk spectroscopy.


\end{widetext}

\end{document}